\documentstyle[12pt,sw20lart]{article}
\input tcilatex
\QQQ{Language}{
British English
}

\begin{document}

\title{Bianchi VIII Empty Futures}
\author{John D. Barrow and Yves Gaspar \\
%EndAName
DAMTP\\
Centre for Mathematical Sciences\\
Cambridge University\\
Wilberforce Road,\\
Cambridge CB3 0WA\\
UK}
\maketitle

\begin{abstract}
Using a qualitative analysis based on the Hamiltonian formalism and the
orthonormal frame representation we investigate whether the chaotic
behaviour which occurs close to the initial singularity is still present in
the far future of Bianchi $VIII$ models. We describe some features of the
vacuum Bianchi $VIII$ models at late times which might be relevant for
studying the nature of the future asymptote of the general vacuum
inhomogeneous solution to the Einstein field equations.
\end{abstract}

\section{Introduction}

Many studies have been made of the asymptotic states of spatially
homogeneous cosmological models. Studies of the early-time behaviour have
focussed upon the possibility of chaotic behaviour in the most general cases 
\cite{BKL}, \cite{M}, \cite{JDB}, while investigations of the late-time
asymptotics have focussed upon the relative generality of cosmological
models which evolve to become at least as isotropic as the observed universe 
\cite{CH}. In this paper we are going to consider some aspects of the
late-time evolution of one of the four most general classes of spatially
homogeneous universe, see \cite{H}, \cite{DLN}, \cite{W}, \cite{Ug}, and 
\cite{BN} for related studies. The Bianchi type $VIII$ universe has not
attracted the same interest from cosmologists as some other homogeneous
models because it does not contain an isotropic Friedmann universe as a
special case. However, it can become arbitrarily close to isotropy. It is
primarily of interest because of its status as one of the most general
homogeneous universes and hence in some sense, yet to be made precise, may
offer a leading order approximation to some part of the general cosmological
solution of Einstein's equations at late times, far from the initial
singularity. In fact, if space is topologically compact it becomes the most
general of the ever-expanding Bianchi type universes \cite{BKod}.

In what follows we will study certain aspects of the late-time behaviour of
vacuum Bianchi type $VIII$ models. In particular, we will try to determine
whether the chaotic Mixmaster behaviour which takes place close to the
initial singularity still takes place in the far future as $t\rightarrow
+\infty $. We will show how a qualitative analysis based on the Hamiltonian
formalism can reveal some basic features of the future asymptote of vacuum
Bianchi $VIII$ models. These provide a basis for studying the nature of the
late-time behaviour of the general solution to the vacuum Einstein field
equations.

It is also of interest to note that late-time cosmological behaviour in
general relativity corresponds to early times in the so-called Pre Big Bang
scenario suggested by symmetries of superstring theory. It has been argued 
\cite{K} that plane-wave space-times are the generic initial states for
these models. So in this context the study of the asymptotic future
behaviour of general vacuum space-times is also of interest. Also, recent
work \cite{DH} has pointed out that inhomogeneous models of low-energy
bosonic field equations for all superstring models and M-theory ( $D=11$
supergravity ) exhibit an oscillatory approach to the initial singularity,
so it would be interesting as well to study how these oscillatory
singularities transform in type $VIII$ universes at late times using a
similar qualitative analysis, see also \cite{BM}.

We will first introduce the field equations for the vacuum Bianchi type $%
VIII $ model and then discuss the orthonormal frame approach and the
Hamiltonian formalism to analyse the possible future asymptotic states.

\section{The field equations for Bianchi type $VIII$}

In the metric approach, where the metric components are the basic variables
of the gravitational field, we can introduce group invariant and
time-independent frame vectors ${e}_a$ such that the line element is given
by 
\[
{ds}^2=-{dt}^2+{{g_{ab}}}{W^a}{W^b}
\]
where $W^a$ are time-independent 1-forms dual to the frame vectors $e_a$. In
the case of vacuum Bianchi type $VIII$ we can introduce three time-dependent
expansion scale factors as 
\[
{{g_{ab}}}=diag(a^2,b^2,c^2)
\]
and Einstein's field equations become \cite{DLN} 
\begin{equation}
{\frac{{{({\dot{a}}{b}{c})^{.}}}}{{abc}}}={\frac{{1}}{{2{a^2}{b^2}{c^2}}}}[{{%
(b^2+c^2)^2}-a^4}]
\end{equation}
\begin{equation}
{\frac{{(a{\dot{b}}c)^{.}}}{{abc}}}={\frac{{1}}{{2{a^2}{b^2}{c^2}}}}[{%
(a^2+c^2)^2}-b^4]
\end{equation}
\begin{equation}
{\frac{{(ab{\dot{c}})^{.}}}{{abc}}}={\frac{{1}}{{2{a^2}{b^2}{c^2}}}}{[{%
(a^2-b^2)^2}-c^4]}
\end{equation}
\begin{equation}
{{\frac{{\ddot{a}}}{{a}}}+{\frac{{\ddot{b}}}{{b}}}+{\frac{{\ddot{c}}}{{c}}}}%
=0
\end{equation}
Here, the dot denotes differentiation with respect to the synchronous time $t
$. If we replace the synchronous time $t$ by a variable $\eta $ defined by 
\[
dt=(abc)d{\eta }
\]
then the field equations (1)-(4) become 
\begin{equation}
2({\ln {a}})^{^{\prime \prime }}={(b^2+c^2)^2}-a^4
\end{equation}
\begin{equation}
2({\ln {b}})^{^{\prime \prime }}={(a^2+c^2)^2}-b^4
\end{equation}
\begin{equation}
2({\ln {c}})^{^{\prime \prime }}={(a^2-b^2)^2}-c^4
\end{equation}
\begin{equation}
({\ln {a^2}})^{^{\prime \prime }}+({\ln {b^2}})^{^{\prime \prime }}+({\ln {%
c^2}})^{^{\prime \prime }}={({\ln {a^2}})^{\prime }}{({\ln {b^2}})^{\prime }}%
+{({\ln {b^2}})^{\prime }}{({\ln {c^2}})^{\prime }}+{({\ln {c^2}})^{\prime }}%
{({\ln {a^2}})^{\prime }}
\end{equation}
where {'} denotes differentiation with respect to $\eta $.

These equations are invariant under the transformation$\ \eta \rightarrow -{%
\eta }$. Now, as is widely exploited in the literature, \cite{M}, \cite{BKL}%
, to a good approximation we have $abc\sim t$ as $t\rightarrow 0,$ which
implies 
\[
\eta \propto {\ln }{(t)}
\]
and the invariance of the field equations (5)-(8) under$\ \eta \rightarrow -{%
\eta }$ corresponds to invariance under 
\[
t\rightarrow {\frac{{1}}{{t}}}
\]
in (1)-(4): an apparent duality transformation relating early and late-time
behaviour. If this symmetry were exact it would imply that the presence of
chaotic behaviour at early times ($t\rightarrow 0$) means that there must
also be chaotic behaviour as $t\rightarrow +\infty ,$ as suggested in ref. 
\cite{H}. A crucial point in the above argument is the accuracy of the
assumption that 
\[
abc\sim t
\]
One can use the field equations (1)-(3) to test this assumption, by looking
at the second derivative of the metric volume element $abc$. Taking the sum
of the equations (1)-(3) we get 
\begin{equation}
{2(abc)}{(abc)^{..}}={(a^2-b^2)^2}+2{a^2}{c^2}+2{c^2}{b^2}+{c^4.}
\end{equation}
Since the right-hand side is the sum of four positive terms, if $abc$ $\neq 0
$ we have 
\begin{equation}
(abc)^{..}>0
\end{equation}

This shows that the relation $abc\propto t$ cannot be exact and so the
duality symmetry argument cannot be applied to relate late and early time
behaviour for the vacuum Bianchi type $VIII$ model. However, the inequality
(10) will be useful in deducing the late-time behaviour as $t\rightarrow
+\infty $. There is another consequence of (10) worth pointing out : if $abc$
is a power law

\[
abc\sim t^n 
\]
then (10) implies $n>1$. Since $dt=(abc){d{\eta }}$, we would have

\[
{\eta }\propto {\frac{{1}}{{1-n}}}{t^{1-n},}\hspace{1.0in}n\neq 1 
\]
Note that near $n=1$ there is a structural instability with respect to the
variation of $t$ with respect to $\eta $. When $n>1$ we have $t\rightarrow 0$
as ${\eta }\rightarrow -\infty $ and $t\rightarrow +\infty $ when ${\eta }%
\rightarrow 0$. However, this differs form the case where $n=1,$ where ${%
\eta }\propto {\ln }(t)$ (which is a good approximation close to the initial
singularity) since here ${\eta }\rightarrow +\infty $ as $t\rightarrow
+\infty $. If $n$ $<1$, then we would have $\eta \rightarrow 0$ as $%
t\rightarrow 0$ and ${\eta }\rightarrow +\infty $ as $t\rightarrow +\infty .$
Tiny variations on either side of $n=1$ would make $\eta $ a bad choice of
variable for an early-time analysis of Mixmaster behaviour. This may be of
importance for numerical studies of Mixmaster behaviour \cite{hob}, which
are notoriously difficult to carry out because of the slow time variation of
the oscillations compared with the growth in volume.

Let $u^a$ be the 4-velocity of the Hubble flow, the Hubble parameter is
defined by 
\[
H={\frac{{1}}{{3}}}{{u}^a}_{;a} 
\]

This parameter can be expressed as follows

\[
H={\ }\frac{\dot l}l 
\]
where $l(t)$ corresponds to a geometric-mean expansion scale factor. The
deceleration parameter $q$ is defined by

\[
q=\ -\frac{l\ddot{l}}{\dot{l}^2} 
\]
The Hubble parameter then satisfies the following equation

\begin{equation}
{\frac{{dH}}{{dt}}}=-(1+q){H^2}
\end{equation}

Defining the a dimensionless shear parameter by

\[
{\Sigma }^2={\frac{{{\sigma }^2}}{{3{H^2}}}} 
\]
with ${{\sigma }^2}={\frac{{1}}{{2}}}{{\sigma }_{ab}}{{\sigma }^{ab}}$,
where ${\sigma }_{ab}$ is the shear tensor, we have the following simple
relation between the shear parameter and the deceleration parameter in
vacuum:

\begin{equation}
q=2{{\Sigma }^2.}
\end{equation}
Equations (11) and (12) can now be used to express the Hubble parameter as 
\begin{equation}
H(t)={\frac{{1}}{{t+2{\int {{{\Sigma }^2}(t)}dt}}}}
\end{equation}
For the isotropic (${\Sigma }=0$) Milne universe we have $H(t)={\ t}^{-1}$
so equation (13) expresses the departure from the Milne expansion rate for
vacuum homogeneous cosmologies.

We specialise to the Bianchi $VIII$ universe and calculate the Hubble scalar
defined above for ${u^a}=(1,0,0,0)$ using the Christoffel symbols for
Bianchi type $VIII$ models \cite{RS}, so:

\begin{equation}
H(t)={\frac{{1}}{{3}}}{\frac{{d(abc)}}{{dt}}}{\frac{{1}}{{abc}}}
\end{equation}
Equation (13) then gives the following expression for $abc$ :

\begin{equation}
abc={\exp ({\int {{\frac{{3}}{{t+2s(t)}}}dt}})}
\end{equation}
with $s(t)={\int {{{\Sigma }^2}dt}}$

From the above equation we can get ${(abc)^{..}}$, as

\begin{equation}
{(abc)^{..}}={\exp ({\int {{\frac{{3}}{{t+2s(t)}}}dt}})}{\frac{{1}}{{{%
(t+2s(t))}^2}}}[9-3(1+2{{\Sigma }^2})]
\end{equation}
For this to agree with the inequality (10) we require

\[
9-3(1+2{{\Sigma }^2})>0 
\]
which implies

\begin{equation}
{\Sigma }^2<1
\end{equation}
As we will see this is consistent with bounds on ${\Sigma }$ given by
equation (28) of the next section.

\section{The orthonormal frame formalism}

We use the orthonormal frame formalism of Ellis and MacCallum \cite{EM}. The
orthonormal frame $e_a$ is chosen to be invariant under the action of the
isometry group with ${e_0}=u,$ so that the commutation functions ${{\gamma }%
^c}_{ab}$ defined by

\[
\lbrack {e_a},{e_b}]={{{\gamma }^c}_{ab}}{e_c} 
\]
depend on the synchronous time $t$ and are the basic variables of the
gravitational field. With this choice of frame, the line-element takes the
form

\[
{ds}^2=-{{dt}^2}+{{\delta }_{ab}}{{\omega }^a}(t){{\omega }^b}(t), 
\]
where ${\omega }^a$ are time dependent 1-forms dual to the vector fields ${e}%
_a$.

The physical state of the models can be described using the components of
the shear tensor ${\sigma }_{ab}$ and the curvature anisotropies ${n}_{ab}$
which occur in the expression of the commutation functions, see Wainwright
and Ellis \cite{WE} and \cite{B}, \cite{BN}. In the above frame we define 
\[
{\sigma }_{ab}=diag({\sigma }_{11},{\sigma }_{22},{\sigma }_{33}) 
\]
and 
\[
n_{ab}=diag(n_1,n_2,n_3) 
\]
and the linear combinations 
\[
{\sigma }_{+}={\frac{{1}}{{2}}}({{\sigma }_{22}}+{{\sigma }_{33}}) 
\]
\[
{\sigma }_{-}={\frac{{1}}{{2{\sqrt{3}}}}}({{\sigma }_{22}}-{{\sigma }_{33}}) 
\]
\[
{n}_{+}={\frac{{1}}{{2}}}({n_2}+{n_3}) 
\]
\[
{n}_{-}={\frac{{1}}{{2{\sqrt{3}}}}}({n_2}-{n_3}) 
\]
Now scale out the overall expansion of the universe by introducing the
expansion-normalised variables

\begin{equation}
{{\Sigma }_{\pm }}={\frac{{{\sigma }_{\pm }}}{{H}}}
\end{equation}
\begin{equation}
{{N}_a}={\frac{{{n}_a}}{{H}}}
\end{equation}
The cosmological model is then fully described by the dimensionless
variables $({\Sigma }_{+},{\Sigma }_{-},N_1,N_2,N_3)$. The variables ${%
\Sigma }_{\pm }$ describe the anisotropy of the Hubble flow and the $N_a$
describe the spatial curvature of the group orbits. If we define a
dimensionless curvature parameter by 
\[
K=-{\frac{{{R}^{(3)}}}{{6{H^2}}}} 
\]
then we can deduce from the generalised Friedmann equation the relation 
\begin{equation}
1={{\Sigma }^2}+K
\end{equation}
Note that the curvature parameter $K$ is related to the variables $N_a$ by

\[
K={\frac{{1}}{{12}}}[{{N_{1}}^2}+{{N_{2}}^2}+{{N_{3}}^2}-2({N_{1}}{N_{2}}+{%
N_{2}}{N_{3}}+{N_{3}}{N_{1}})] 
\]

We can introduce a dimensionless time variable $\tau $ defined by

\begin{equation}
{\frac{{dt}}{{d{\tau }}}}={\frac{{1}}{{H}}}
\end{equation}
and related to the mean expansion scale factor, $l(t)$, by

\begin{equation}
l=l_0{e^\tau }
\end{equation}
These variables yield the following simplified evolution equations for the
vacuum type $VIII$ state variables :

\begin{equation}
{{\Sigma}_{\pm}}^{^{\prime}}=-(2-q){{\Sigma}_{\pm}}-{S}_{\pm}
\end{equation}

\begin{equation}
{N_{1}}^{^{\prime}}=(q-4{{\Sigma}_{+}}){N_{1}}
\end{equation}

\begin{equation}
{N_{2}}^{^{\prime}}=(q+{2{\Sigma}_{+}}+2{\sqrt{3}}{{\Sigma}_{-}}){N_{2}}
\end{equation}

\begin{equation}
{N_3}^{^{\prime }}=(q+2{{\Sigma }_{+}}-2{\sqrt{3}}{{\Sigma }_{-}}){N_3}
\end{equation}
where $^{\prime }$ denotes derivatives with respect to $\tau $ and the
functions ${S}_{\pm }$ are given by :

\[
{S_{+}}={\frac{{1}}{{6}}}[{({N_{2}}-{N_{3}})^2}-{N_{1}}(2{N_{1}}-{N_{2}}-{%
N_{3}})] 
\]

\[
{S_{-}}={\frac{{1}}{{2{\sqrt{3}}}}}({N_{2}}-{N_{3}})({N_{1}}-{N_{2}}-{N_{3}}%
) 
\]

This system, together with equation (20), is an autonomous system of
first-order ordinary differential equations for which a monotone function
can be found with respect to the time variable $\tau $. Different Bianchi
types lead to different restrictions on the ranges of the variables $N_a$.
For Bianchi type $VIII$ we require $N_1<0$, $N_2>0$, $N_3>0$ and the set
defined by these constraints is an invariant set for the evolution
equations. On this invariant Bianchi type $VIII$ set there exists the
following monotone increasing function : 
\begin{equation}
Z=({N_1}{N_2}{N_3})^2
\end{equation}
As a consequence of eq. (20) the following bounds can placed on the state
variables \cite{WE} :

\begin{equation}
0 \le {{\Sigma}^2} \le 1
\end{equation}
\begin{equation}
-1 \le {{\Sigma}_{\pm}} \le +1
\end{equation}
\begin{equation}
0 \le {N_{1}}^2 \le 12
\end{equation}
\begin{equation}
0 \le {N_{1}}({N_{2}}-{N_{3}}) \le 12
\end{equation}
\begin{equation}
0 \le -2{{N_{1}}}({N_{2}}+{N_{3}}) \le 12
\end{equation}
\begin{equation}
0 \le {({N_{2}}-{N_{3}})^2} \le 12
\end{equation}

Therefore, the invariant set $S$ for Bianchi type $VIII$ is unbounded in
such way that ${N_2}+{N_3}$ can assume arbitrarily large values. Together
with the existence of the monotone increasing function $Z,$ given by
equation (27), this means we do not expect Bianchi type $VIII$ orbits to
approach an equilibrium point as $\tau \rightarrow +\infty $.

\section{Hamiltonian Cosmology}

As discussed in section 2, in the metric approach, we can introduce a set of
group-invariant and time-independent one-forms ${W}^a$, so that with an
arbitrary time variable $t^{\prime }$ the metric has the form

\begin{equation}
{{ds}^2}=-{N(t^{\prime }){dt^{\prime }}^2}+{{g_{ab}(t^{\prime })}}{W^a}{W^b}
\end{equation}
In vacuum, the Bianchi type $VIII$ metric is diagonal and we write the
metric components in the form \cite{M} 
\begin{equation}
{{g_{ab}}}=diag({e^{2{\beta }_1}},{e^{2{\beta }_2}},{e^{2{\beta }_3}})
\end{equation}
with 
\[
{\beta }_1={{\beta }^0}-2{{\beta }^{+}} 
\]
\[
{\beta }_2={{\beta }^0}+{{\beta }^{+}}+{\sqrt{3}}{{\beta }^{-}} 
\]
\[
{\beta }_3={{\beta }^0}+{{\beta }^{+}}-{\sqrt{3}}{{\beta }^{-}} 
\]
In the Hamiltonian approach the spatial Ricci curvature defines a potential
function $V({\beta }_{+},{\beta }_{-})$, 
\[
V({\beta }_{+},{\beta }_{-})=-{e^{-2{\beta }^0}}{R^{(3)}} 
\]
The variable ${\beta }^0$ is related to the overall length scale for the
metric (34) and to the Hubble parameter by the relations

\begin{equation}
l=l_0{e^{{\beta }^0}}
\end{equation}

\begin{equation}
H={N^{-1}}{\dot{{\beta }^0}}
\end{equation}
The evolution of the Bianchi type $VIII$ universe is then described by the
motion of a point in the mini-superspace $({\beta }^0,{\beta }^{+},{\beta }%
^{-})$ evolving in the potential $V({\beta }^{+},{\beta }^{-})$. The
variable ${\beta }^0$ can be chosen to be the time variable so that $\dot{{%
\beta }^0}=1$ and $N={H}^{-1}$. Equations (22) and (36) imply that we have ${%
{\beta }^0}={\tau }$.

This Hamiltonian approach allows us to study the early and late-time
asymptotic behaviour of Class A Bianchi-type universes qualitatively. For
vacuum Bianchi type $I$, the potential function vanishes, $V({{\beta }^{\pm }%
})=0$, so that the evolution is a straight line in ${\beta }^{\pm }$-space.
For Bianchi type $VIII$ the equipotential curves of $V({{\beta }^{\pm }})$
are approximately triangular in shape, \cite{RS}, with an infinite open
channel which starts at the vertex on the ${\beta }^{+}$ axis and becomes
increasingly narrow as ${{\beta }_{+}}\rightarrow +\infty $. As ${\tau }%
\rightarrow +\infty $ (evolution away from the initial singularity) the
walls of the triangle contract. If the universe point moves initially in a
straight line (corresponding to Kasner-like behaviour) it will bounce off a
potential wall and then move along another straight line (a transition to
another Kasner-like behaviour). The triangular shape of the potential allows
the evolution to be chaotic, which corresponds to the Mixmaster behaviour
and is familiar from studies of the evolution near the singularity ($\tau
\rightarrow -\infty ).$ But, what happens as $\tau \rightarrow +\infty $ ?
The Hamiltonian picture implies that the triangular potential contracts, so
this would enable the universe point to bounce off the walls in a chaotic
sequence, ad infinitum. We shall argue however from the inequality (10), or
the bound (28), and the fact that ${N_2}+{N_3}$ can take arbitrarily large
values, that this chaotic future evolution cannot occur as $\tau \rightarrow
+\infty $, contrary to the claim of Halpern, \cite{H}.

In order to show this, we introduce two further equations, relating the ${%
\Sigma }_{\pm }$ to the ${\beta }^{\pm }$ variables. Since \cite{WE}

\begin{equation}
{\Sigma }_{\pm }={\frac{{d{{\beta }^{\pm }}}}{{d{\tau }}},}
\end{equation}
where ${\Sigma }_{\pm }$ coincides with expression calculated in the
orthonormal frame formalism, using equation (12), we have the relation

\begin{equation}
{\frac{{q}}{{2}}}={\Sigma}^2={{{\Sigma}_{+}}^2}+{{{\Sigma}_{-}}^2}=({\frac{{d%
{{\beta}^{+}}}}{{d{\tau}}}})^2+({\frac{{d{\beta}^{-}}}{{d{\tau}}}})^2
\end{equation}

The Bianchi type $VIII$ model exhibits Mixmaster behaviour towards the past
and this chaotic evolution is reflected in the infinite sequence of Kasner
era's and bounces off the triangular potential walls in the Hamiltonian
picture \cite{BKL}, \cite{M}, \cite{RS}. When bounces occur, the function

\begin{equation}
{({\frac{{{d{\beta }^{+}}}}{{d{\tau }}}})^2}+{({\frac{{d{\beta }^{-}}}{{d{%
\tau }}}})^2}
\end{equation}
exhibits a series of peaks (a bounce produces an abrupt change in ${\beta }%
^{+}$ or ${\beta }^{-}$). By equation (39) this implies that ${{\Sigma }^2}$
should exhibit a series of peaks as well. A graph of the function ${\Sigma }$
is given in Wainwright and Ellis \cite{WE}. Now, if the triangular potential
contracts as $\tau \rightarrow +\infty $, the interval between bounces
decreases as well. This means that time derivatives of ${\beta }_{\pm }$
could diverge (the peaks of ${{\beta }_{\pm }}(\tau )$ would become
increasingly narrow) and so the function (40) would be unbounded. This would
imply, by equation (39), that ${{\Sigma }^2}$ would be unbounded, which
contradicts the inequality (17) or (28). In this picture of evolution in an
eternally contracting potential, one can imagine two solutions which avoid
this problem.

First, it could be that ${{\beta }_{\pm }}(\tau )$ is of the form

\[
{{\beta }_{\pm }}(\tau )={{e^{-a{\tau }}}osc{(b{e^{a{\tau }}})}}+c, 
\]
with $a$, $b$, $c$ constants. In the above the $osc$ function could have a
periodic or chaotic oscillatory behaviour. Clearly this could give
increasingly rapid chaotic oscillations of decreasing amplitude and using
equation (38) one finds a bounded, eternally chaotic, oscillating ${{\Sigma }%
_{\pm }}$. However, defining the function ${p_0}(\tau )$ by 
\[
{\frac{{d{p_0}}}{{d{\tau }}}}=(2-q){p_0} 
\]
one can show the following relations in the Hamiltonian approach \cite{WE}

\begin{equation}
{{\tilde N}_a}={\frac{{12{e^{2{{\beta }_a}}}}}{{(-{p_0})}}}
\end{equation}
where ${{\tilde N}_a}={n_a}{N_a,}$ ( no sum ) with ${n_1}={{C^1}_{23}}$, ${%
n_2}={{C^2}_{31}}$ and ${n_3}={{C^3}_{12}.}$ Using equations (41), (35) and
(12), one finds, 
\begin{equation}
|{N_1}{N_2}{N_3}|\sim {\exp {(6{\tau }-6{\ \int {(1-{{\Sigma }^2}){d{\tau }}}%
})}}
\end{equation}
We see that if ${{\Sigma }_{\pm }}$ oscillates eternally and satisfies (17)
or (28), then (42) implies that $|{N_1}{N_2}{N_3}|$ increases as $\tau
\rightarrow +\infty $ (see end of section 3). Now, equation (30) implies
that $N_1$ is bounded. Since ${\beta }_{\pm }$ tend to constant values, the
expression for ${\beta }_1$ in (35) and equation (41) with $a=1$ implies
that 
\[
{\frac{{e^{2{{\beta }_0}}}}{{(-{p_0})}}} 
\]
is bounded, and this means that equations (35) and (41) with $a=2$ and $a=3$
will imply that $N_2$ and $N_3$ are each bounded. Hence the Bianchi type $%
VIII$ invariant set would be bounded and this clearly contradicts the fact
that ${N_2}+{N_3}$ diverges as $\tau \rightarrow +\infty $ because of
equation (42).

Second, the universe point could slow down, but then we could end up
reaching a point where ${{\beta }_{\pm }}\rightarrow $ constant and ${{%
\Sigma }^2}\rightarrow 0$. One can easily see that in this case equation
(23) will imply that equation (20) will not be satisfied. Indeed, ${{\Sigma }%
_{-}}=0$ will imply that $N_2=N_{3\text{ }}$ for Bianchi $VIII$ , so
equation (23) requires ${{{\Sigma }_{-}}^{^{\prime }}}=0$. Equations
(24),..., (26), (41) and (35) imply that the $N_a$ become constants, but
then (23) requires ${{S}_{+}}=0$, otherwise ${{\Sigma }_{+}}$ does not
satisfy the bound (17) or (28). In order to have ${S_{+}}=0$ we must satisfy
either $N_1=0$ or $N_1=N_2=N_3$. The first case implies that the curvature
parameter $K=0$, and this is inconsistent with the Friedmann equation (20),
and the second possibility implies that $N_1=N_2=N_3=0$ because of the sign
restrictions mentioned in the previous section for Bianchi type $VIII$. This
also leads to $K=0$ and contradicts (20) as well. Note that this argument is
not necessarily true in the non-vacuum situation (see appendix).

So what are the alternative possibilities consistent with the Hamiltonian
picture? The infinitely long channel along the ${\beta }^{+}$ axis offers a
solution. The universe point could evolve eternally between the two walls of
the infinite channel, for instance there could be an infinity of
oscillations between the two walls of the channel. This evolution would
correspond to a non-chaotic oscillation of the universe point about the
axisymmetric solution (see below) characterised by

\[
{\beta}^{-}=0 
\]

In this case, because the channel becomes increasingly narrow as ${\beta }%
_{+}\rightarrow +\infty $, we expect ${\beta }_{-}$ to undergo increasingly
rapid oscillations of decreasing amplitude as $\tau \rightarrow +\infty $.
Similar behaviour of the universe point when interacting with a channel of
the Bianchi IX potential can be found in Ryan and Shepley \cite{RS} for a
so-called 'mixing bounce'. Equation (42) also implies that $%
N_2+N_3\rightarrow +\infty $ as $\tau \rightarrow +\infty $ and so the
orbits in the state space escape to infinity.

Non-chaotic oscillatory behaviour for Bianchi type $VIII$ seems also to be
observed in the behaviour of the Weyl curvature of other Bianchi type
universes according to Coley \cite{C} and in Wainwright et al. \cite{WHU}.
In this interesting work the late-time behaviour of Bianchi type $VII_0$
models with perfect fluid source was investigated, (see also refs. \cite{CH}%
, \cite{DLN}, \cite{BS}. If the universe contains matter with pressure $p$
and density $\rho ,$ obeying an equation of state

\[
p=(\gamma -1)\rho 
\]
with $\gamma $ constant, then if ${\gamma }<{\frac{{4}}{{3}}}$ the universe
isotropizes with respect to shear but not with respect to the Weyl
curvature. So even if the shear decreases at late times, the anisotropic
Weyl curvature dominates the dynamics, producing a form of self-similarity
breaking at late times. If ${\frac{{4}}{{3}}}<{\gamma }\leq 2$, then ${{%
\Sigma }^2}$ is oscillatory as $\tau \rightarrow +\infty ,$ with ${\beta }%
_{-}$ a decreasing oscillatory function, while the Weyl scalar diverges. It
was further argued that since an open subset of a more general class of
models (i.e. Bianchi type $VIII$) will display features of Bianchi $VII_0$
models, the non-self-similar behaviour of the Weyl curvature could occur at
some stage during the evolution of Bianchi $VIII$ models.

Let us introduce the dimensionless electric and magnetic parts of the Weyl
curvature

\[
{\tilde{E}}_{ab}={E_{ab}}{\frac{{1}}{{H^2}}} 
\]
\[
{\tilde{H}}_{ab}={H_{ab}}{\frac{{1}}{{H^2}}} 
\]
For all Bianchi class A models, in the standard frame ${\tilde{E}}_{ab}$ and 
${\tilde{H}}_{ab}$ are diagonal and we define:

\begin{equation}
{\tilde{E}}_{+}={\frac{{1}}{{2}}}({\tilde{E}}_{22}+{\tilde{E}}_{33})
\end{equation}

\begin{equation}
{\tilde{E}}_{-}={\frac{{1}}{{2{\sqrt{3}}}}}({\tilde{E}}_{22}-{\tilde{E}}%
_{33})
\end{equation}
and likewise for ${\tilde{H}}_{+}$ and ${\tilde{H}}_{-}$. By defining 
\begin{equation}
{N}_{+}={\frac{{1}}{{2}}}(N_2+N_3)
\end{equation}

\begin{equation}
{N}_{-}={\frac{{1}}{{2}}}(N_2-N_3).
\end{equation}
We have the following useful relations between the state variables
introduced in the orthonormal frame approach and the electric and magnetic
parts of the Weyl curvature :

\begin{equation}
{\tilde{E}}_{+}={\Sigma}_{+} + ( {{{\Sigma}_{+}}^2}-{{{\Sigma}_{-}}^2})+{%
S_{+}}
\end{equation}

\begin{equation}
{\tilde{E}}_{-}={\Sigma}_{-} - 2{{\Sigma}_{+}}{{\Sigma}_{-}}+{S_{-}}
\end{equation}

\begin{equation}
{\tilde{H}}_{+}=-{\frac{{3}}{{2}}}{{N_{1}}}{{\Sigma}_{+}} -3{N_{-}}{{\Sigma}%
_{-}}
\end{equation}

\begin{equation}
{\tilde{H}}_{-}=-3{N_{-}}{{\Sigma }_{+}}+{\frac{{1}}{{2}}}({N_1}-4{{N_{+}}}){%
{\Sigma }_{-}.}
\end{equation}
If ${N_2}+{N_3}$ becomes arbitrarily large, then by equation (50), we can
have 
\begin{equation}
\lim_{\tau \rightarrow +\infty }{{\tilde{H}}_{-}}=+\infty
\end{equation}
and the Weyl scalar, given by

\[
W={{{\tilde{E}}_{+}}^2}+{{{\tilde{E}}_{-}}^2}+{{{\tilde{H}}_{+}}^2}+{{{%
\tilde{H}}_{-}}^2,} 
\]
can become arbitrarily large as $\tau \rightarrow +\infty $, see Wainwright 
\cite{W} for a discussion of the self-similarity breaking. Similar behaviour
for the Weyl curvature was found in perfect fluid Bianchi type $VII_0$
universes with equation of state parameter satisfying $1<\gamma \leq 2$ \cite
{WHU}.

\section{ The axisymmetric (LRS) vacuum Bianchi type $VIII$ model}

Our qualitative analysis shows that the late-time behaviour of vacuum
Bianchi type $VIII$ models will be non-chaotic. Evolution could occur which
exhibits oscillations about the axisymmetric (LRS) vacuum Bianchi type $VIII$
model with ${{\Sigma }_{-}}=0$. The asymptotic late-time behaviour of this
solution is known \cite{private}. The Bianchi type $VIII$ LRS model is
described by the following restrictions on the state variables, and these
define an invariant set for the evolution equations (23)-(26) :

\[
N_1<0 
\]
\[
N_2=N_3>0 
\]
\[
{\Sigma }_{-}=0 
\]
This state space is unbounded, since $N_2$ can take arbitrarily large
values. However, we can use the following bounds on ${N_1}{N_2}$ (see
inequalities (28)-(33)):

\[
0\le -{N_1}{N_2}\le 3 
\]
We introduce a new variable $Z$ defined by Wainwright, \cite{private},

\[
{Z}^2=-{N_1}{N_2} 
\]
It can then be shown that the system of evolution equations (23)-(26) admits
the following equilibrium point :

\[
N_1=0 
\]
\[
{\Sigma }_{+}={\frac{{1}}{{2}}} 
\]
\[
Z={\frac{{3}}{{2}}} 
\]
This corresponds to a flat plane-wave asymptote, which has $W=0$. More
precisely, it is the Bianchi type $III$ form of flat space-time, with metric

\begin{equation}
{ds}^2=-{dt}^2+{t^2}({{dx}^2}+{e^{2x}}{{dy}^2})+{{dz}^2.}
\end{equation}
The Bianchi type $VI_h$ diagonal plane-wave metric with $h=-1$ reduces to
this line element for a particular choice of parameters.

Since the above equilibrium point solution, (52), is of LRS Bianchi type $%
III $ with ${{n^\alpha }_\alpha }=0$, it is a particular case of the
following general metric which includes Bianchi types $I$, $VI_0$, $VIII$, $%
V $ and $VI_h$ with ${{n^\alpha }_\alpha }=0$ \cite{EM}

\begin{eqnarray}
{ds}^2 &=&-{{dt}^2}+(X^2+{Y^2}{w^2}){{dx}^2}-2{Y^2}w{e^{-({a_0}+{q_0})x}}{dx}%
{dy}  \nonumber \\
&&+{Y^2}{e^{-2({a_0}+{q_0})x}}{{dy}^2}+{Z^2}{e^{2({a_0}-{q_0})x}}{{dz}^2}
\end{eqnarray}
where $a_0$ and $q_0$ constants and $X(t)$, $Y(t)$, $Z(t)$, $w(t)$ are
functions of the synchronous time $t,$ and 
\[
w(t)=-2b{\ \int {{X}{Y^{-3}}{Z^{-1}}{dt}}} 
\]
where $b$ is constant. The line element (52) has,

\[
b=0,X=Y=t,Z=1,{a_0}={q_0}=-{\frac{{1}}{{2}}.} 
\]
Now since this line element is a future attractor for the LRS vacuum Bianchi
type $VIII,$ and since (53) contains non-LRS solutions, it is interesting to
perform a stability analysis of this line element for non-LRS perturbation
modes. The perturbation technique for special cases of (53) has already been
developed by Barrow and Sonoda \cite{BS}. In this approach one defines a new
time variable as

\[
{\frac{{d}}{{du}}}={X(t)}{\frac{{d}}{{dt}}}=^{\prime } 
\]
as well as the new variables $F={\frac{{X^{^{\prime }}}}{{X}}}$, $G={\frac{{%
Y^{^{\prime }}}}{{Y}}}$ and $H={\frac{{Z^{^{\prime }}}}{{Z}}}$ so that the
field equations for (53) in vacuum become

\begin{equation}
{{F^{^{\prime }}}}=2({{a_0}^2}+{{q_0}^2})-{\frac{{2{b^2}{X^2}}}{{{Y^4}{Z^2}}}%
}-{F}(G+H),
\end{equation}
\begin{equation}
{{G^{^{\prime }}}}=2({{a_0}^2}+{a_0}{q_0})+{\frac{{2{b^2}{X^2}}}{{{Y^4}{Z^2}}%
}}-G(G+H),
\end{equation}
\begin{equation}
{{H^{^{\prime }}}}=2({{a_0}^2}-{a_0}{q_0})-H(G+H),
\end{equation}
with the first integral constraints

\begin{equation}
{\frac{{2{b^2}{X^2}}}{{{Y^4}{Z^2}}}}=2F(G+H)+2GH-(6{{a_0}^2}+2{{q_0}^2})
\end{equation}
and 
\begin{equation}
2{F}{a_0}={q_0}(G-H)+G+H.
\end{equation}

The unperturbed solution is the line element (52) with ${F_0}=1$, ${G_0}=1$
and ${H_0}=0$. We require equations for the following small perturbations, $%
f,g,h$ :

\[
F={F_0}+f, 
\]
\[
G={G_0}+g, 
\]
\[
H={H_0}+h. 
\]
Using the equations (54)-(58), one gets the following (the equation for $F$
is eliminated using the constraint (58)) :

\begin{equation}
{{g^{^{\prime}}}}=-4g-4h
\end{equation}

\begin{equation}
{{h^{^{\prime}}}}=-h
\end{equation}

The eigenvalues associated with this problem are ${{\lambda }_1}=-4$ and ${{%
\lambda }_2}=-1$ and so the solution (52) is asymptotically stable under
small non-LRS perturbations consistent with (53), thus supporting the
prediction that the universe point will evolve about the axisymmetric
solution with ${\beta }_{-}=0$ in the narrow infinite channel along the ${%
\beta }_{+}$-axis. Taking into account the results of the previous section,
this means we expect the future asymptotic state of the full non-LRS vacuum
Bianchi type $VIII$ to have

\[
{\Sigma }={\frac{{1}}{{2}}.} 
\]

The fact that ${\Sigma }_{-}$ vanishes is a consequence of the following :
since ${{\beta }_{-}}\rightarrow 0$, the line element (34) becomes
axisymmetric and so the field equations (5), (6) and (7) lead to the LRS
solution for the scale factors, thus ${{\beta }_{+}}\rightarrow {\frac{{\tau 
}}{{2}}}$, however from equation (41) one finds that if ${\Sigma }_{-}$
oscillates and does not decrease, then the curvature parameter $K$ can be
unbounded, which contradicts the generalised Friedmann equation (20). Also,
although ${\Sigma}_{-}$ decreases, it can be shown to undergo increasingly
rapid oscillations : if ${\beta}_{-}$ is bounded, oscillatory and
decreasing, then equations (23), (38) and (41) imply that ${\beta}_{-}$ (
and thus ${\Sigma}_{-}$ ) undergoes increasingly rapid oscillations. If the
universe point oscillates about the axisymmetric solution, then the metric
shares a feature of some gravitational-wave space-times, namely that the
solution oscillates about a given form of flat space-time as $\tau
\rightarrow +\infty $.

\section{Conclusion}

A qualitative analysis based on the Hamiltonian formalism and the
orthonormal frame approach gives the following information about the future
asymptotic state of vacuum Bianchi type $VIII$ models :

$a$. The chaotic Mixmaster solution close to the initial singularity evolves
at late times into a simpler non-chaotic solution, contrary to the claims of
ref. \cite{H}, and corresponding to the fact that the universe point evolves
along the infinite open channel of the Bianchi $VIII$ potential and
oscillates about the LRS Bianchi type $VIII$ solution in such a way that the
shear parameter exhibits increasingly rapid oscillations as $\tau
\rightarrow +\infty $ .

$b.$The form of the line element (34) tends to the Bianchi $III$ form of
flat space-time because ${\beta }_{-}\rightarrow 0$. It is interesting to
compare this with asymptotic states of other vacuum Bianchi types : Bianchi
types $IV$, $VI_h$, and $VII_h$ tend to plane-wave space-times; Bianchi type 
$VII_0$ tends to the special equilibrium point ${N_1}=0$, $N_2=N_3=$
constant, with ${{\Sigma }_{+}}=-1$ and ${{\Sigma }_{-}}=0$, which in the
state space is a line emanating from a point corresponding to the Taub form
of flat space; finally, Bianchi type $V$ tends to the Milne form of flat
space. So it appears that different forms of flat and plane-wave space-times
play an important role in the future state of these eternally expanding
vacuum Bianchi models and one might expect that they occur in a
leading-order approximation to part of the general inhomogeneous vacuum
solution to the Einstein field equations at late times.

\section{Acknowledgements}

We would like to thank John Wainwright for sending us helpful information
about the late-time behaviour of LRS Bianchi type $VIII$ models.

\section{Appendix}

For a non-vacuum homogenous cosmology the Friedmann equation (20) becomes

\begin{equation}
1={{\Sigma }^2}+K+{\Omega }
\end{equation}
where $\Omega $ is the density parameter defined by ($8\pi G=1$) 
\[
{\Omega }={\frac{{\rho }}{{3H^2}}} 
\]
with $\rho $ the energy density of a perfect fluid with equation of state $%
p=({\gamma }-1){\rho .}$ It appears that if $0<\gamma \le {\frac{{2}}{{3}}}$%
, the LRS Bianchi $VIII$ models admits the flat Friedmann model as future
attractor i.e. ${\Sigma }=N_1=N_2=N_3=0$ and ${\Omega }=1$, so that the
argument developed in section 4 does not hold. This is simply the well known
cosmic no hair attractor of power-law inflationary universes whose matter
content violates the strong energy condition. In fact, for non-vacuum
Bianchi type $VIII$, the analysis of section 4 can be adapted as follows
when the strong energy condition holds. Note first that the deceleration
parameter is now given by 
\[
q=2{{\Sigma }^2}+{\frac{{1}}{{2}}}(3{\gamma }-2){\Omega } 
\]
and so we have $q\geq 0$ if $3{\gamma }-2\geq 0$. For all $\gamma $
satisfying this inequality, one can apply the analysis of section 4 and the
universe point will again evolve in the infinite channel as $\tau
\rightarrow +\infty $.

We end this appendix by a remark about the stiff fluid case with ${\gamma }%
=2 $. This case is of interest namely because the stiff fluid is equivalent
to the presence of a massless scalar field. The evolution equation for $%
\Omega $ is

\begin{equation}
{\frac{{d{\Omega }}}{{d{\tau }}}}=[2q-(3{\gamma }-2)]{\Omega }
\end{equation}
Now when $\gamma =2$, the field equations (1),..., (3) are identical for
vacuum and non-vacuum cases. So, the inequality (17) is valid and this
corresponds to 
\[
q<2 
\]
and so equation (62) implies that ${\Omega }\rightarrow 0$ as $\tau
\rightarrow +\infty $, so that the evolution tends to a vacuum solution.

\end{document}